\documentclass[twocolumn]{aastex6}
\usepackage[utf8]{inputenc}
\usepackage{natbib}
\usepackage{verbatim}
\usepackage{hyperref}
\usepackage{subcaption}
\usepackage{footmisc}

\date{July 31, 2019 (Published)}
\begin{document}

\title{Hyper Wide Field Imaging of the Local Group Dwarf Irregular Galaxy IC 1613: An Extended Component of Metal-poor Stars}

\author{Ragadeepika Pucha\altaffilmark{1}, Jeffrey L. Carlin\altaffilmark{2}, Beth Willman\altaffilmark{1,3}, Jay Strader\altaffilmark{4}, David J. Sand\altaffilmark{1}, Keith Bechtol\altaffilmark{2,5}, Jean P. Brodie\altaffilmark{6}, Denija Crnojevi\'c\altaffilmark{7}, Duncan A. Forbes\altaffilmark{8}, Christopher Garling\altaffilmark{9}, Jonathan Hargis\altaffilmark{10}, Annika H. G. Peter\altaffilmark{11}, Aaron J. Romanowsky\altaffilmark{6,12}}

\altaffiltext{1}{Steward Observatory, The University of Arizona, 933 North Cherry Avenue, Tucson, AZ 85721, USA} \email{rpucha@email.arizona.edu}
\altaffiltext{2}{Large Synoptic Survey Telescope, 950 North Cherry Avenue, Tucson, AZ 85719, USA}
\altaffiltext{3}{Association of Universities for Research in Astronomy, 950 North Cherry Avenue, Tucson, AZ 85719, USA}
\altaffiltext{4}{Department of Physics and Astronomy, Michigan State University, East Lansing, MI 48824, USA}
\altaffiltext{5}{Department of Physics, University of Wisconsin, Madison, WI 53706, USA}
\altaffiltext{6}{University of California Observatories, 1156 High Street, Santa Cruz, CA 95064, USA}
\altaffiltext{7}{University of Tampa, 401 West Kennedy Boulevard, Tampa, FL 33606, USA}
\altaffiltext{8}{Centre for Astrophysics and Supercomputing, Swinburne University, Hawthorn VIC 3122, Australia}
\altaffiltext{9}{CCAPP and Department of Astronomy, The Ohio State University, Columbus, OH 43210, USA}
\altaffiltext{10}{Space Telescope Science Institute, 3700 San Martin Drive, Baltimore, MD 21218, USA}
\altaffiltext{11}{CCAPP, Department of Physics, and Department of Astronomy, The Ohio State University, Columbus, OH 43210, USA}
\altaffiltext{12}{Department of Physics \& Astronomy, San Jos\'e State University, One Washington Square, San Jose, CA 95192, USA}

\begin{abstract}
Stellar halos offer fossil evidence for hierarchical structure formation. Since halo assembly is predicted to be scale-free, stellar halos around low-mass galaxies constrain properties such as star formation in the accreted subhalos and the formation of dwarf galaxies. However, few observational searches for stellar halos in dwarfs exist. Here we present {\it gi} photometry of resolved stars in isolated Local Group dwarf irregular galaxy IC 1613 ($M_{\star} \sim 10^8 M_{\odot})$. These Subaru/Hyper Suprime-Cam observations are the widest and deepest of IC 1613 to date. We measure surface density profiles of young main-sequence, intermediate to old red giant branch, and ancient horizontal branch stars outside of $12\arcmin$ ($\sim 2.6$ kpc; 2.5 half-light radii) from the IC 1613 center. All of the populations extend to $\sim 24\arcmin$ ($\sim 5.2$ kpc; 5 half-light radii), with the older populations best fit by a broken exponential in these outer regions. Comparison with earlier studies sensitive to IC 1613's inner regions shows that the density of old stellar populations steepens substantially with distance from the center; we trace the $g$-band effective surface brightness to an extremely faint limit of $\sim 33.7$ mag arcsec$^{-2}$. Conversely, the distribution of younger stars follows a single, shallow exponential profile in the outer regions, demonstrating different formation channels for the younger and older components of IC 1613. The outermost, intermediate-age and old stars have properties consistent with those expected for accreted stellar halos, though future observational and theoretical work is needed to definitively distinguish this scenario from other possibilities.
\end{abstract}

\section{Introduction} \label{Intro}

The outskirts of galaxies are important testing grounds for galaxy formation and evolution. The currently accepted $\Lambda$CDM model of hierarchical structure formation implies that dark matter halos grow via mergers and the accretion of smaller halos \citep[][and references therein]{White+1978, Springel+2006}. Such interactions leave imprints on the outer regions of galaxies. The long dynamical timescales in these outer regions preserve the fossil evidence of their formation. Studying the outskirts of a galaxy can therefore reveal its formation history  \citep[e.g.,][]{Bullock+2005}. 

Extensive observational and theoretical work shows that the two nearest massive galaxies, the Milky Way and M31, have distinct halo properties that reflect the diversity of assembly histories expected in $\Lambda$CDM.

The outskirts of the Milky Way reveal an almost spherical distribution of old population stars with large random motions. This ``stellar halo'' constitutes just $\sim 1$\% of the total stellar mass of the galaxy, but has provided many clues to the Milky Way's formation (see \citealt{Helmi2008} review).  \citet{Searle+1978} suggested that the halo is built-up from the remnants of disrupted dwarf galaxies.  The stellar halo is also predicted to include stars that were formed {\it in situ} (i.e, in the inner regions of the galaxies) but got displaced to the halo because of mergers \citep{Zolotov+2009}.  Recent surveys confirm these predictions in the form of many complex structures and tidal streams in the stellar halo of the Galaxy, as well as a more diffuse halo component \citep[e.g.,][]{Belokurov+2013, Shipp+2018}. The evidence for hierarchical assembly of M31 is nearly as extensive, with many coherent streams and other tidal structures \citep{McConnachie2009}, though, notably, M31's stellar halo is more metal-rich and massive than that of the Milky Way (e.g., \citealt{Ibata+2014}).

This work on the Milky Way and M31, as well as voluminous evidence in more distant galaxies (e.g., \citealt{Martinez+2010,Crnojevic+2016,Monachesi+2016}) has confirmed  that dwarf galaxies are indeed the building blocks of the stellar halos of massive galaxies.

Since structure formation in $\Lambda$CDM is scale invariant, dwarf galaxies themselves should also form by hierarchical merging and hence might also have stellar halos. The main uncertainty is the star formation efficiency in the subhalos that merged to make the dwarfs: star formation in such halos is both less efficient and more stochastic than at Milky Way--like masses (e.g., \citealt{Garrison+2017}). Hence it might be the case that dwarfs do possess stellar halos, but with a lower stellar mass fraction and a high level of scatter among dwarfs.

There is a long history of observational work relevant to stellar halos in dwarfs, even if not phrased in such terms. For example, structures more extended than the main galactic body have been found in many dwarf galaxies. In 1963, Baade found that the young blue population in the nearby dwarf galaxy IC 1613 is embedded in a more extended red population.  Defined as the Baade sheet \citep{Sandage1971}, such extended components were later found in many dwarf galaxies, including WLM \citep{Minniti+1996}, Phoenix \citep{Martinez-Delgado+1999, Hidalgo+2009}, DDO 187 \citep{Aparicio+2000}, DDO 190 \citep{Aparicio2000}, Leo A \citep{Vansevicius+2004}, and NGC 3109 \citep{Hidalgo+2008}. 

\begin{figure*}[t!]
	\centering
    \includegraphics[width = 0.6\textwidth]{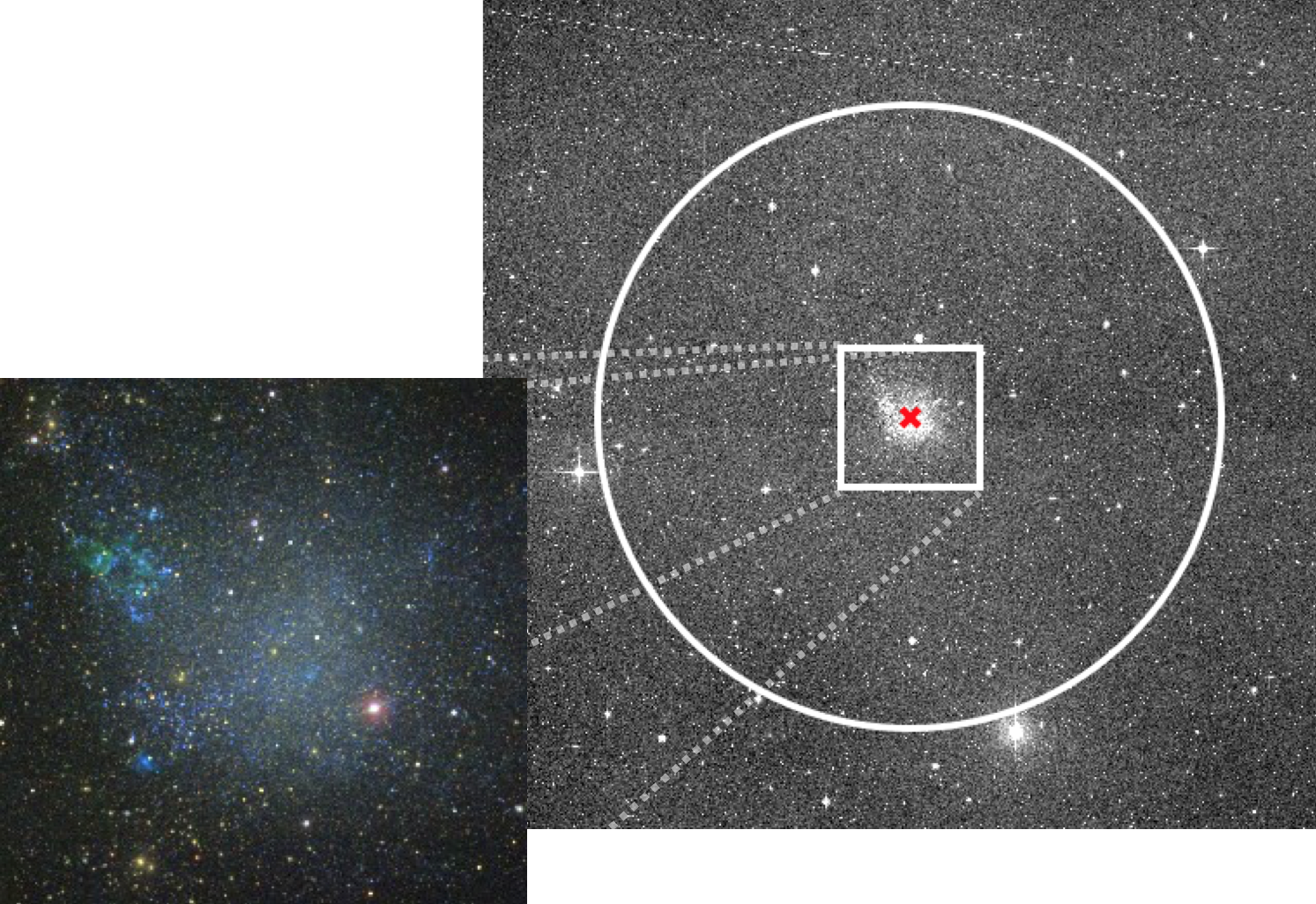}
    \caption {DSS image ($2^{\circ}$ $\times$ $2^{\circ}$ in size) centered on IC~1613 with the white circle showing the $1^{\circ}.5$ diameter field of view of our HSC observations. The center of the field, $(\alpha, \delta) = (01^{h} 04^{m} 47^{s}.80, +02^{\circ} 07' 04''.00)$, is marked. \textit{Inset:} Color composite of IC 1613 from SDSS images ({\it image credits: \url{www.legacysurvey.org}}).}
\label{fig:CenterField_Inset}
\end{figure*}

Perhaps the most extensive dataset is for the periphery of Magellanic Clouds, which also reveal extended stellar envelopes \citep{Nidever+2011, Nidever+2018}. The structure of these outer populations is highly distorted due to the interactions of the Clouds with the Milky Way and with each other. The origin of the well-established extended envelope of the Large Magellanic Cloud is uncertain: it could be due to tidal stripping of the outer disk, or to ``classical" accretion of dwarfs. The recent discovery of a substantial population of low-mass dwarfs associated with the Large Magellanic Cloud \citep{Koposov+2015, Drlica-Wagner+2016} strengthens the latter scenario.
The Clouds are not the only dwarfs with known satellites: dwarf galaxy companions have been found around various Magellanic analogs (\citealt{Martinez-Delgado+2012}; MADCASH Survey: \citealt{Carlin+2016}). A reasonable expectation is that a subset of these satellites will be tidally destroyed and accreted by their host halos, providing additional 
motivation to search for accreted stellar halos in dwarf galaxies.

Stellar halos of galaxies are predicted and observed to be of very low surface brightness. This makes it challenging to study their presence quantitatively through diffuse light. Instead, the most tractable way to search for stellar halos is through deep, wide-field imaging now available on a number of large telescopes located at excellent sites.

The isolated Local Group dwarf irregular IC~1613 is a good candidate to search for a stellar halo.
It is a gas-rich, low-luminosity galaxy with an absolute magnitude of $M_{V} = -15.2$ \citep{McConnachie2012}, corresponding to a stellar mass of $\sim 10^8 M_{\odot}$. Its high Galactic latitude implies low foreground extinction, while it is near enough ($725\pm17$ kpc; \citealt{Hatt+2017}) that its bright stars are accessible in a reasonable exposure time from the ground. IC~1613 
has been continuously forming stars throughout its lifetime, and hosts stars ranging from [Fe/H] $\sim$ --2 to --0.8 \citep{Cole+1999, Skillman+2003, Bernard+2007, Weisz+2014, Skillman+2014}. A variety of studies sensitive to different stellar populations at different radii have found that the young stars are more centrally concentrated than the intermediate-age and old stars  \citep{Cole+1999, Albert+2000, Borissova+2000, Skillman+2003, Sibbons+2015, McQuinn+2017}; though, there has been no definitive detection of a stellar halo as distinct from an extension of the primary body of IC~1613.

The aim of this paper is to map IC~1613 with 
Subaru/Hyper Suprime-Cam (HSC) over a wider area and to fainter magnitudes than previous studies, generating the most sensitive map of its outer structure to date. 

The paper is organized as follows. The observations and information about how the data were reduced is presented in Section~\ref{DataReduction}. Section~\ref{Analysis} describes the method of selecting different stellar tracers from the galaxy and goes on to explain the various analysis done with them. Section~\ref{Results} compares our results with past works and places this project in a theoretical context. The conclusions are summarized in Section~\ref{Conclusions}. 

\section{Observations and Data Reduction} \label{DataReduction}

\begin{figure}[t!]
	\centering
    \includegraphics[width = 0.9\columnwidth]{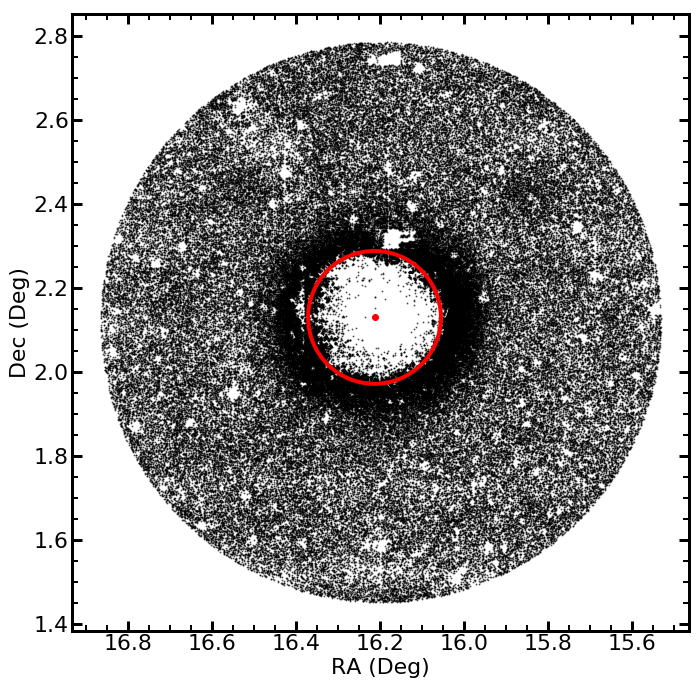}
    \caption{Spatial distribution of all point sources from the HSC field. The center is marked by the red point as calculated in Section~\ref{EllipseFit}. The central region is crowded and so few sources are detected, resulting in a nearly empty region. The red circle with a projected radius of 2~kpc (at the distance of IC~1613) provides a sense of scale. Empty spots on the periphery are due to bright foreground stars.}
    \label{fig:ScatterPlot}  
\end{figure}

IC~1613  was observed on 2015 October 16, with the HSC \citep{Miyazaki+2012} on the Subaru Telescope.  The large aperture (8.2 m) of the telescope, along with the $1^\circ.5$ field of view of the HSC, makes it a good fit to study faint resolved stellar populations over the wide-field study of IC~1613. A single field was observed with exposure times of 10 $\times$ 300~s in {\it g} (known as ``HSC-G'' in Subaru parlance) and 10 $\times$ 120~s in {\it i} (``HSC-I''), with $5\times30$ s exposures taken in each band to recover bright stars that saturate in the long exposures. The seeing varied between $\sim0''.5$ and $1''.0$. Figure~\ref{fig:CenterField_Inset} shows the observed field overlaid on a DSS image of IC 1613. At the assumed distance of 725 kpc, our observations extend to a projected distance of $\sim$ 8.4 kpc from the center of the galaxy. Given the half-light radius ($r_h$) of $4\arcmin.8\pm0\arcmin.3$ ($\sim$ 1.01 kpc, \citealt{Hunter+2006}), these data extend to about 8.3 $r_h$ in projected radius. They also reach a limit of about 4 mag below the tip of the red giant branch, making this the widest and deepest study of IC 1613 to date (see Figures~\ref{fig:ScatterPlot} and~\ref{fig:CMD}). 

The images were processed using a prototype version of the LSST optical imaging data processing pipeline. A fork of the LSST stack was used for HSC data processing (\emph{hscPipe}, \url{http://hsc.mtk.nao.ac.jp/pipedoc_e/index.html}), but the HSC-specific tools have since been merged back into the LSST Stack \citep{Bosch+2017,Huang+2018}. For overview information on LSST and its Data Management system, see \citet{Ivezic+2008} and \citet{Juric+2015}.
In brief, bias subtraction, dark current correction, and flat-fielding using dome flats were applied to the images, followed by nonlinearity, brighter-fatter, and cross-talk corrections. The individual short and long exposures were coadded, and photometry was performed on a per-band basis. The stack produces many photometric outputs. As discussed in the next subsection, two are relevant for this paper: forced point spread function (PSF)-fitting photometry (appropriate for stars) and \emph{cmodel} photometry (a linear combination of exponential and $r^{1/4}$ models, appropriate for galaxies). The images were then astrometrically and photometrically calibrated against Pan-STARRS1 Processing Version 2 \citep{Schlafly+2012,Tonry+2012}. All the astrometric and photometric measurements presented in this work are thus in the Pan-STARRS1 system. The extinction values for the sources in each band are computed using the dust map from \citet{Schlegel+1998} and the extinction coefficients from \citet{Schlafly+2011}, assuming an extinction to reddening ratio ($R_{V}$) of 3.1. The mean extinction across the field is $E(B-V) \sim 0.023$.

\begin{figure}[h!]
	\centering
    \includegraphics[width = 0.95\columnwidth]{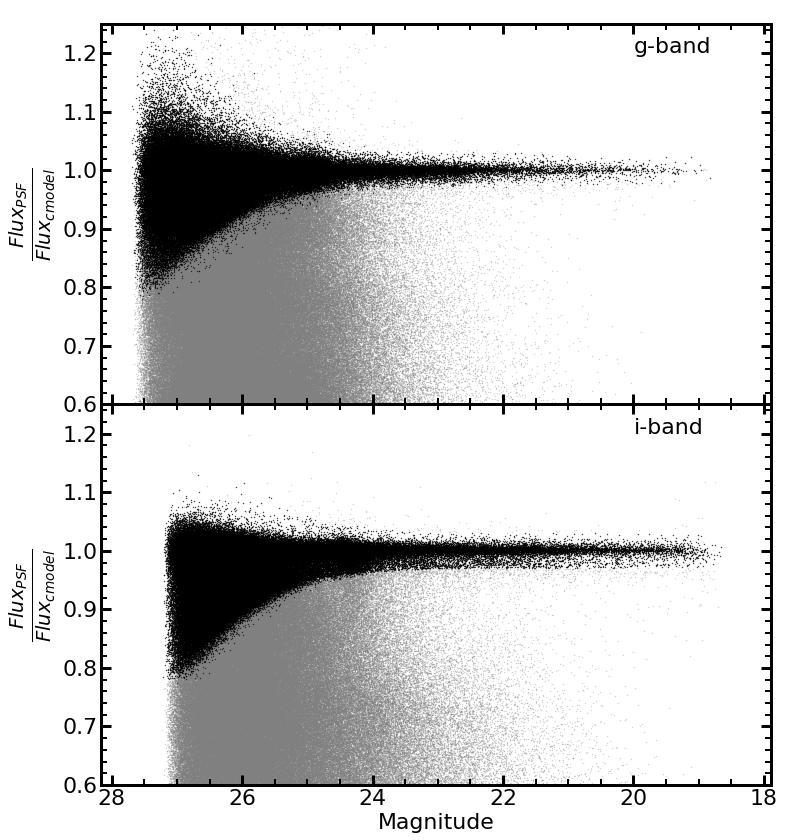}
    \caption{Star--galaxy separation: Here we show the ratio of PSF flux to {\it cmodel} flux as a function of magnitude ($g$-band on the top panel, $i$-band on the bottom panel). Stars have an expected value of unity. We classify sources within $1\sigma$ of this value as point sources (black; 165,550 sources), compared to the full set of sources (gray; 580,166 sources), most of which are background galaxies.}
 	\label{fig:StarGalaxySep}
\end{figure} 

\begin{figure}[t!]
	\centering
 	\includegraphics[width = 0.9\columnwidth]{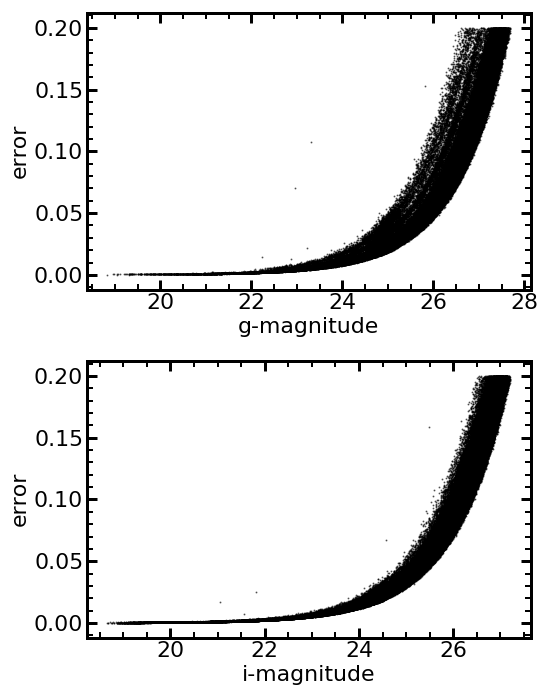}
    \caption{Uncertainties as a function of magnitude in both $g$ (upper panel) and $i$ (lower panel). The 5$\sigma$ point-source depths are $g\sim27.0$ mag and $i\sim26.5$ mag.}
    \label{fig:ErrorPlot}
\end{figure}

\subsection{Star--Galaxy Separation} \label{SG_Separation}

The resulting catalog has $g$ and $i$ photometry for an enormous number of sources within the HSC field of view, but only a fraction of these sources are resolved stars associated with IC 1613. Many of the sources, especially toward faint magnitudes, are barely resolved (or unresolved) distant background galaxies.
A central challenge in resolved stellar populations is separating the stars of interest from these background galaxies.
 
As a simple but effective means of star/galaxy separation, we use the ratio of PSF to \emph{cmodel} flux. True stars scatter around a ratio of unity, while galaxies show fainter PSF fluxes than their true fluxes and so are found at lower values.
This statistic has been successfully used for star/galaxy separation with a similar dataset by \citet{Carlin+2017}.

Figure~\ref{fig:StarGalaxySep} plots the PSF to \emph{cmodel} flux ratio as a function of magnitude, showing that stars are well-separated from galaxies at $g \lesssim 24.5$ and $i \lesssim 24$, but less so at fainter magnitudes. Hence some galaxy contamination is inevitable at the faintest magnitudes. To balance this contamination with a reasonable sample of stars, we classify sources as stars if they are within $1\sigma$ of the expected stellar flux ratio of unity (where $\sigma$ is the per-object uncertainty on the PSF/\emph{cmodel} flux ratio). There is an asymmetric extension toward $f_{\rm PSF}/f_{cmodel} < 1$ at bright ($i < 23$) magnitudes. We have confirmed that most of the sources in this feature are RGB stars in the main body of IC~1613, suggesting that the feature is likely due to the elevated background in the crowded inner regions. We also only consider sources detected at $5\sigma$ or above, corresponding to a depth of $g \sim 27$ and $i \sim 26.5$ in the uncrowded regions of the image (Figure~\ref{fig:ErrorPlot}).

Figure~\ref{fig:ScatterPlot} shows the spatial distribution of the point sources selected in this manner. A red circle of radius 2 kpc (9\arcmin.5) is overplotted on the scatter plot to visually set the scale. The galaxy center, as calculated in Section~\ref{Analysis}, is also marked. It is clear that crowding affects the photometry in these inner regions of the galaxy where few stars are detected. In the next subsection we discuss artificial star tests to determine the regions in which the photometry is reliable.  We also note that several stellar overdensities are apparent in the IC~1613 field (e.g. at R.A. = $15^\circ.85$, Dec = $+2^\circ.4$). Inspection of the color-magnitude diagrams (CMDs) corresponding to these overdensities do not reveal any obvious stellar population that could be associated with IC~1613 or, for instance, a very faint dwarf galaxy at the distance of IC~1613.  A detailed, quantitative search for dwarf galaxies associated with IC~1613 will be presented in a future publication.

\begin{figure}[t!]
	\centering
    \includegraphics[width = 0.95\columnwidth]{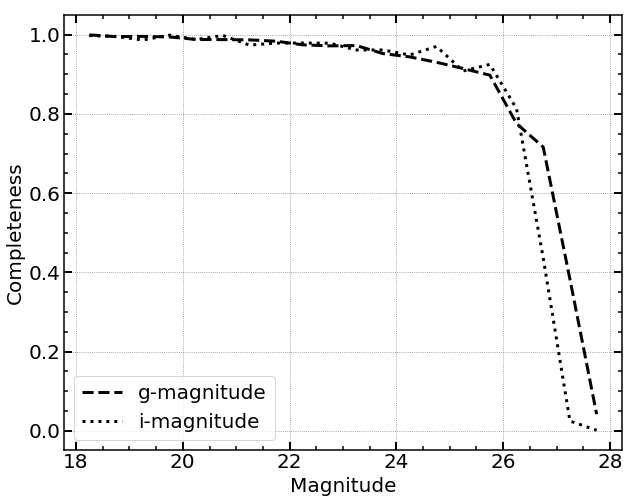}
	\caption{Completeness as a function of magnitude obtained from the artificial star tests. The data are more than 80\% complete for magnitudes brighter than 26 mag in both $g$ and $i$ bands.}
    \label{fig:Completeness_Mag}
\end{figure}

\subsection{Artificial Star Tests} \label{Completeness}

We used artificial star tests to quantify and correct for the incompleteness in point sources as a function of magnitude and position. Fake stars were added into the processed images using the {\it Synpipe} software \citep{Huang+2018}. The input catalog contained stars with magnitudes $16 < i < 28$, linearly weighted so that there are $\sim3$~times as many objects at $i=28$ than at $i=16$, and uniformly distributed between $-1 < g-i < 2.5$. A total of 2000 artificial stars were inserted per $\sim11\arcmin.2 \times 11\arcmin.2$ patch (i.e., $\sim16$ stars ~arcmin$^{-2}$). These fake stars were added into images using the measured PSF and noise characteristics of each individual image, and the resulting images were processed in the same way as for the real observations.

\begin{figure}[t!]
    \centering
    \includegraphics[width = 0.95\columnwidth]{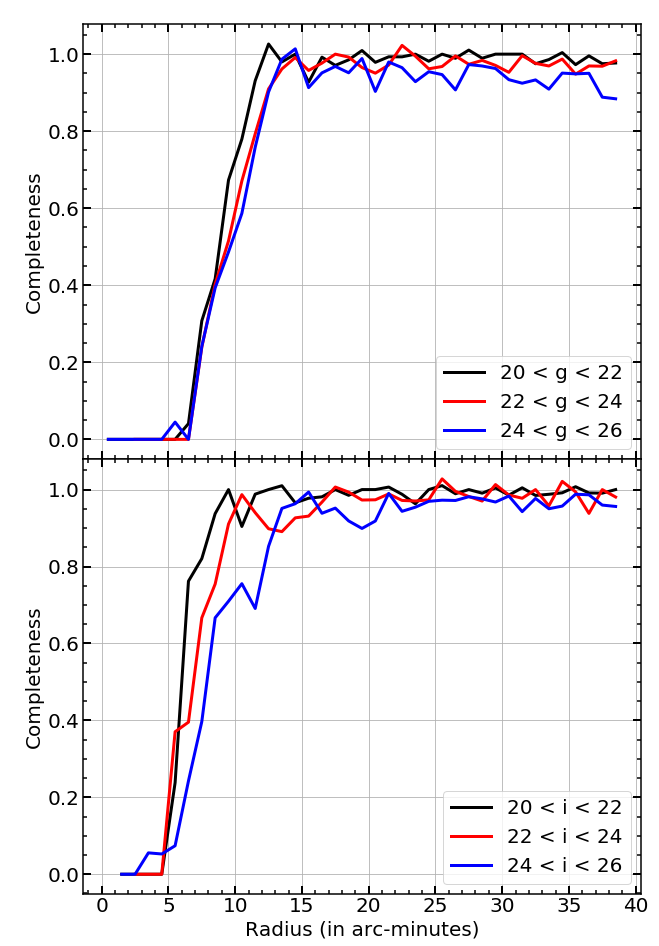}
    \caption{Completeness as a function of radius for different magnitude bins (Top: $g$, Bottom: $i$). The photometry is 80\% complete for the positions lying outer to $\sim$ 12$'$ from the center of the galaxy for both the bands.}
    \label{fig:Completeness_Pos}
\end{figure}

Figure~\ref{fig:Completeness_Mag} shows the completeness as a function of magnitude for both {\it g} and {\it i} bands. Taking bins of 0.5 mag in each filter, the completeness was computed as the number of output fake stars in each bin divided by number of input fake stars in the same bin. The photometry is more than 80\% complete for magnitudes brighter than 26 mag in both filters. 
Due to crowding, completeness also depends on the distance from the galaxy center in a magnitude-dependent manner. Completeness as a function of position and magnitude is  shown in Figure~\ref{fig:Completeness_Pos}. The data are almost 80\% complete for both bands for regions that lie outside $\sim$ $12\arcmin$ from the center of IC 1613. 
Hence we exclude this region ($< 12\arcmin$) that 
is strongly affected by crowding from our analysis.
 
\section{Analysis} \label{Analysis}
\subsection{The Color--Magnitude Diagram} \label{CMD}

\begin{figure*}[t!]
	\centering
    \includegraphics[width = 0.9\textwidth]{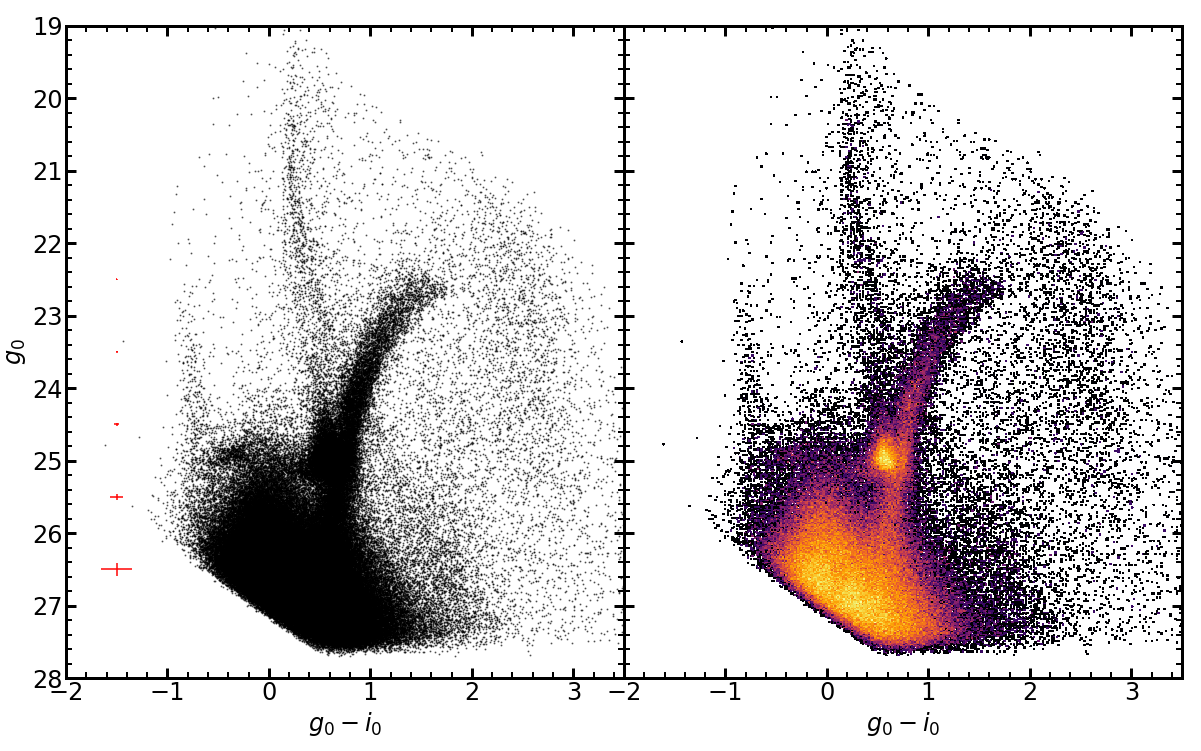}
    \caption{Left Panel: Color--magnitude diagram of point sources from $12\arcmin$--$40\arcmin$ (2.5--8.4 kpc) of the center of IC~1613. The magnitudes have been corrected for extinction. The median magnitude and color uncertainties are overplotted in the left panel. The different regions of the color--magnitude diagram: red giant branch, main sequence, red clump, and horizontal branch stars are all clearly visible.  The vertical feature brighter than $g \sim 23.5$ is the Sgr stream. Right Panel: Hess diagram of point sources over the same region.}
    \label{fig:CMD}
\end{figure*}

Figure~\ref{fig:CMD} shows the color-magnitude diagram of point sources within the radial range of $12\arcmin$--$40\arcmin$ (2.5--8.4 kpc) from the center of IC~1613. First we discuss the features not associated with IC~1613: well-measured point sources extend as faint as $g \sim$ 27 mag, but a large fraction of the sources fainter than 26 mag are unresolved background galaxies (Figure~\ref{fig:StarGalaxySep}). Except for the upper giant branch of IC~1613, the majority of the points redder than $g-i \gtrsim 1.5$ mag are foreground stars of the Milky Way disk. 

There is also a narrow plume of sources that extends upwards from $g\sim 23$ at $g-i \sim 0.5$. This feature appears to be a well-defined main sequence observed at a narrow range in heliocentric distance. We note that IC~1613 is in the same direction as previous detections of tidal debris from the Sagittarius dwarf spheroidal \citep[e.g.,][]{Koposov+2012}. The distance of the Sagittarius point sources along the line of sight to IC 1613 is $\sim$ 25 kpc \citep[]{Belokurov+2014}. Adopting this distance, and ages and metallicities of the Sagittarius stream found in the literature \citep[e.g.,][]{Marconi+1998, Bellazzini+1999b}, we confirmed the observed feature to be consistent with that expected from Sagittarius. Hence, we conclude that this additional feature in the CMD is indeed debris from the Sagittarius dwarf galaxy along the line of sight to IC~1613.

Focusing on IC~1613 itself, several features that can be clearly attributed to specific stellar populations are  visible in Figure~\ref{fig:CMD}:

\begin{itemize}
\item[---] The red giant branch (RGB), consisting of stars older than 1 Gyr, is the most distinguishing feature in the CMD. It becomes distinct from the blob of unresolved galaxies at $g \gtrsim 25.5$. In this same magnitude range, the photometric errors are smaller than the observed width of the RGB; thus the observed thickness must be intrinsic (see Figure~\ref{fig:CMD}). This is primarily due to a metallicity spread among the old stellar populations in the galaxy (the expected contribution due to a range of distances along the line of sight is $< 0.01$ mag and hence negligible).

\item[---] The red clump is another prominent feature in the CMD, located around $g \sim 25$ and $g-i \sim 0.5$. The red clump occurs for intermediate-age to old stellar populations with ages of 1--10 Gyr, though its position varies little with age in this range.

\item[---] The main sequence (MS) is the roughly vertical feature of blue stars at $-1 \lesssim (g-i) \lesssim -0.5$, which consists of young (less than $\sim 1$ Gyr) stars. While it makes up only a tiny fraction of the identifiable stars in our radial range, its presence does imply  recent star formation in the outskirts of IC~1613.

\item[---] The horizontal branch (HB) stars are from ancient stellar populations (age $\gtrsim 10$ Gyr). The HB---clearly visible in the CMD at $g\sim25$ -- can be separated roughly into two separate features: a red horizontal branch (RHB) extending blueward from the red clump, and the blue horizontal branch (BHB) extending redward from $g-i \sim -0.5$. The apparent gap in the HB is the RR Lyrae instability strip, which is clearly well-populated (see also \citealt{Cole+1999}).

\end{itemize}

\begin{figure}[!t]
	\centering
    \includegraphics[width = 0.95\columnwidth]{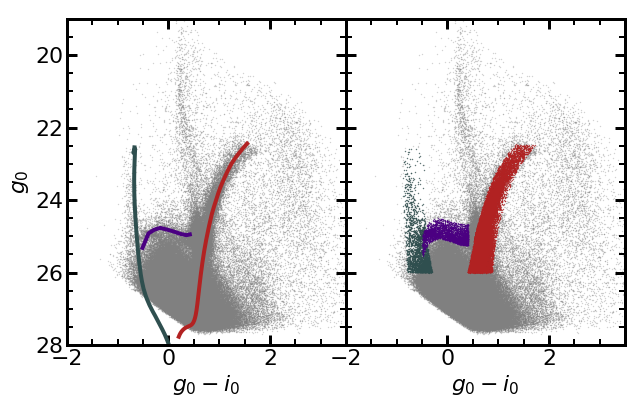}
    \caption{Left Panel: Color--magnitude diagram of point sources within $12\arcmin$--$40\arcmin$ of the center of IC 1613, with overplotted isochrones used for the selection of the tracer stellar populations. The theoretical {\it MIST} isochrones for RGB (age = 10 Gyr and [Fe/H] = $-1.5$) and MS (age = 100 Myr and [Fe/H] = $-0.8$) are shown in red and dark green colors. In addition, the fiducial ridgeline used for the selection of HB ([Fe/H] = $-1.5$) is shown in purple. Right Panel: Selection of stellar tracers. The selected RGB, HB, and MS stars are shown in red, purple, and dark green colors, respectively.}
    \label{fig:StellarTracers}
\end{figure}

We focus our analysis in this paper on three of these tracers -- the young MS stars, intermediate-to-old age RGB stars, and old HB stars. The {\it MIST} \citep{Choi+2016, Dotter2016} isochrones based on ages and metallicities of the populations given in \citet{Cole+1999} and \citet{Skillman+2003} are overplotted on the CMD at the distance adopted for IC 1613. RGB (age = 10 Gyr, [Fe/H] = --1.5) and MS (age = 100 Myr, [Fe/H] = --0.8) isochrones are overplotted on the CMD and provide a good visual match to the observed populations (Figure~\ref{fig:StellarTracers}). In addition, a fiducial ridgeline of an old metal poor ([Fe/H] = --1.5) globular cluster \citep[NGC 5272;][]{Bernard+2014} is overplotted on the CMD for the horizontal branch. This ridgeline provides a reasonable fit to the shape of the HB, and is used in our selection method (see Section~\ref{StellarTracers}) to include the most likely HB stars.

\subsection{Selection of Stellar Tracers} \label{StellarTracers}
The three stellar tracers used in this project---MS, RGB, and HB---trace out young, intermediate to old, and old populations, respectively. The weight of a star's likelihood of membership in the stellar tracer selection is calculated using its separation from the isochrone and its photometric errors as:

\begin{center}
\begin{large}
$w = \exp{(-\frac{(q-q_{iso})^{2}}{2(\sigma^2 + \delta^2)})}$
\end{large}
\end{center}

\noindent where $\delta$ is the intrinsic width of the IC 1613 stellar locus. For the RGB and MS, $q = (g-i)$ is the color of the point source and $q_{iso}$ is the corresponding interpolated point on the isochrone for the same $g$-magnitude. For the HB, $q = g$ is the $g$-magnitude of the point source and $q_{iso}$ is the corresponding interpolated point on the isochrone for the same $g-i$ color. $\sigma$ is the combined photometric error of the point source from both of the filters. 

The stellar tracers are separated by selecting all the points with weight $w \ge $ 0.4. The stars selected for further analysis are shown overplotted on the CMD in the right panel of Figure~\ref{fig:StellarTracers}. Only stars with {\it g} brighter than 26 mag are considered. This ensures using high signal-to-noise data (see Figure~\ref{fig:ErrorPlot}) while also avoiding severe unresolved galaxy contamination at the fainter end (Figure~\ref{fig:StarGalaxySep}).

\subsection{Center and Ellipticity of IC~1613} \label{EllipseFit}

We first focus on determining the overall structure of IC~1613 by finding its center and ellipticity. To calculate the center and ellipticity of IC~1613's outskirts, we repeatedly draw random samples of 15,000 (out of a total 20,274) RGB stars with $r >12'$ without replacement. The center and ellipticity are taken from the best-fit ellipse for the isodensity contour that is 3$\sigma$ above the background level outside $25\arcmin$ from the center of the galaxy (e.g. \citealt{Hammel2018}). By repeating this process for 1000 iterations, we find the coordinates of the center to be J2000 ($\alpha, \delta$) = ($16^{\circ}.21405 \pm 0^{\circ}.00650, 2^{\circ}.12918 \pm 0^{\circ}.00774$). This is formally offset by about $1\arcmin.1 \pm0\arcmin.4$ to the north east of the literature center: ($\alpha, \delta$) = ($16^{\circ}.19917, 2^{\circ}.11778$), from \citet{McConnachie2012}.

\begin{figure}[t!]
	\centering
    \includegraphics[width = 0.99\columnwidth]{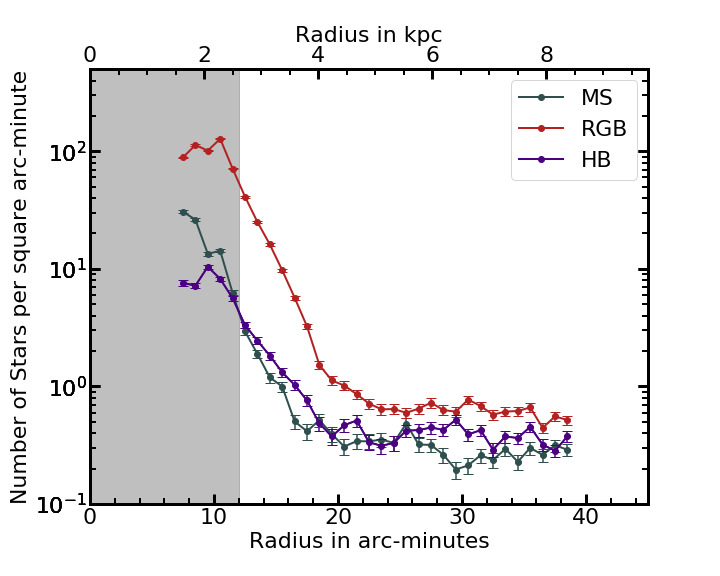}
    \caption{Surface density profiles for different stellar tracers---RGB (red), HB (purple), and MS (dark green)---in circular bins of $1\arcmin$ width centered on IC~1613. The crowding limits our completeness in the shaded gray area. Thus, we consider only the region to the right of this shaded area ($> 12\arcmin$) in our analysis.}
    \label{fig:SurfaceDensityProfiles}
\end{figure}

\begin{figure*}[t!]
    \centering
    \includegraphics[width = 1.0\textwidth]{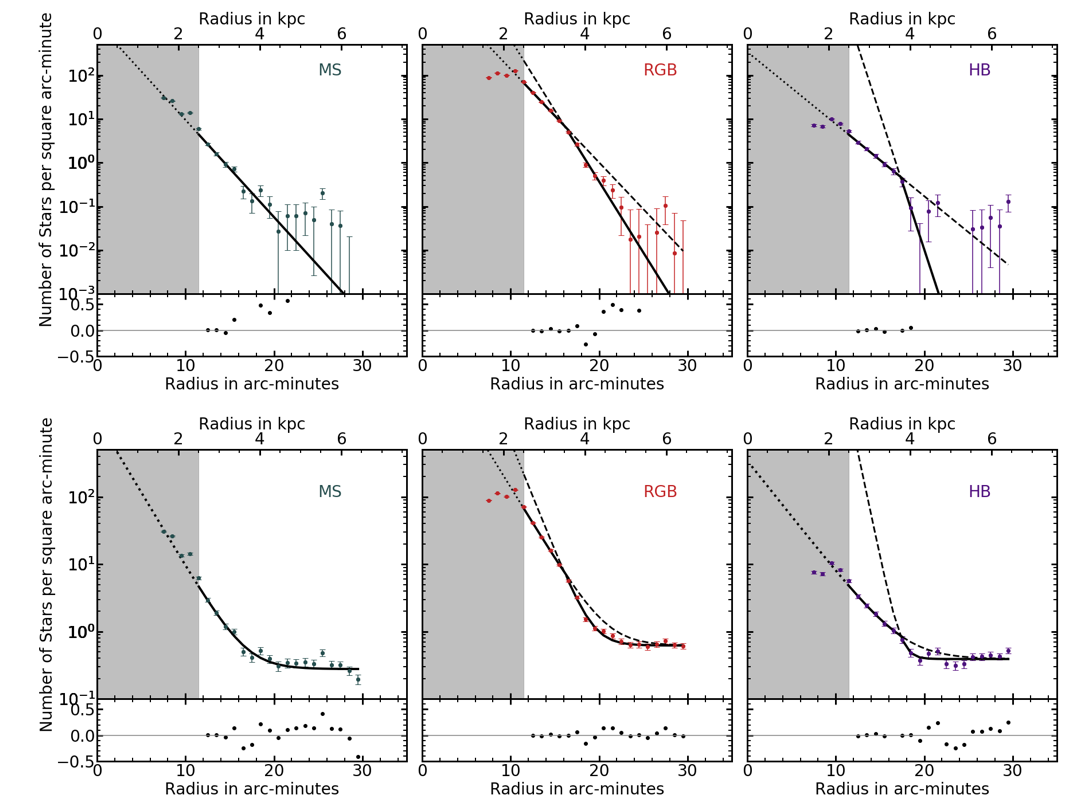}
    \caption{Top: Best fit exponential profiles overplotted on the background subtracted surface density profiles of each of the stellar tracers.  Bottom: Best fit exponential profiles with added background overplotted on the surface density profiles of each of the stellar tracers.  The curves are extrapolated to the central regions and are shown with dotted
    lines. The lower panel of all the six plots show the fractional residuals left over after the fits. Left: MS star profile with an overplotted single exponential fit. Middle: RGB star profile with an overplotted broken exponential, with a break at $\sim 15\arcmin.8$. Right: HB star profile with an overplotted broken exponential, with a break at $\sim 17\arcmin.2$. } 
    \label{fig:ExpFits}
\end{figure*}

By fitting ellipses to different isodensity contour levels, we find that the ellipticity varies minimally from $\sim$ 0.10 at $\sim 13\arcmin$ to $\sim$ 0.04 at $\sim 18\arcmin$, compared to a central ellipticity of $\sim 0.11$ \citep{McConnachie2012}. Our results show that the galaxy has a nearly circular morphology in the outer regions. We assume these values for IC~1613's center and ellipticity for the rest of the paper. 

\subsection{Radial Profiles of Stellar Tracers} \label{SDP}
We use surface density profiles to trace out the radial structure of  IC~1613 for stars with $g < 26$. The profiles for each of three stellar tracers (MS, RGB, and HB) are given in Figure~\ref{fig:SurfaceDensityProfiles}. Since the ellipticity is nearly zero, we use circular projected radial bins of $1\arcmin$ width to calculate the profiles. The surface density is corrected for completeness by applying the radial corrections shown in Figure~\ref{fig:Completeness_Pos}. 

All profiles reach the background at $\sim22\arcmin$--24\arcmin ($\sim$5.2~kpc; $5 r_h$), which is much further than typical studies of old populations in dwarf galaxies, owing to our large field of view. It is interesting that the young MS population extends to almost $\sim$ 20\arcmin. To rule out BHB contamination, we compared the MS profile to an MS selection limited to $g < 25$, and found that it does not change this result. The surface brightness that corresponds to these faint regions is $\sim 33.7$ mag arcsec$^{-2}$ (in $g$), which is calculated at the background surface density level of $\sim$ 0.62 RGB stars per square arc-minute, by assuming a Salpeter IMF \citep{Salpeter1955} and a 10 Gyr population with [Fe/H] = $-1.5$.

\begin{table*}[t!]
    \centering
    \caption{Scale Lengths of Exponentials for the Three Stellar Tracers. {\bf Note:} The inner scale length is from \citet{Bernard+2007}. The rest of the columns are from the fits done in this work. $r_b$ denotes the radius of the break between the two exponentials.}
    \begin{tabular}{|c|c|c|c|c|c|}
    \hline
    {\bf Stellar Tracer} & \vtop{\hbox{\strut{\bf Inner $r_0$ ($r < 12\arcmin$)}} \hbox{\strut{\citep{Bernard+2007}}}} & \vtop{\hbox{\strut {\bf Intermediate $r_0$}}\hbox{\strut ($12\arcmin < r < r_b$)}}  & \vtop{\hbox{\strut {\bf Outer $r_0$}}\hbox{\strut ($r > r_b$)}} & {\bf $r_b$} & $\chi_{red}^2$\\
    \hline
    MS & $1\arcmin.19 \pm 0\arcmin.04$ & $1\arcmin.95 \pm 0\arcmin.13$ & $1\arcmin.95 \pm 0\arcmin.13$ & - & 13.4 \\
    RGB & $3\arcmin.8 \pm 0\arcmin.1$ & $2\arcmin.04 \pm 0\arcmin.05$ & $1\arcmin.34 \pm 0\arcmin.07$ & $15\arcmin.8$ & 4.0 \\
    HB & - & $2\arcmin.63 \pm 0\arcmin.23$ & $0\arcmin.70 \pm 0\arcmin.33$ & $17\arcmin.2$ & 29.9 \\

    \hline
    
    \end{tabular}
    \label{tab:Parameters}
\end{table*}

The gray area in Figure~\ref{fig:SurfaceDensityProfiles} marks the boundary ($\sim 12\arcmin$) beyond which crowding no longer significantly affects our data (see Section \ref{Completeness} and Figure~\ref{fig:Completeness_Pos}). Hence, for further analysis, we use only the data points right of the gray area ($r > 12\arcmin$).

Next we fit functions to the background subtracted radial profiles. The background for each of the stellar tracers is computed as the completeness corrected surface density between $25\arcmin$ and $38\arcmin$. We focus on single exponential and broken exponential fits to the data, because they have provided good descriptions of dwarf galaxies in the past \citep{Hidalgo+2003, Vansevicius+2004, Hunter+2006, Zhang+2012, Herrmann+2013}, and experimenting with such fits suggested they would also provide reasonable fits to the IC 1613 data. The functional form for the exponential fit is: 

\begin{center}
    $\rho(r) = \rho_{0}$  $exp{\Big(-\frac{r}{r_{0}} \Big)}$
\end{center}

\noindent where $\rho_{0}$ is the central surface density and $r_{0}$ is the scale length, which is defined as the radius at which the density falls by $e$.  S{\'e}rsic profiles also fit the broken exponential profiles (e.g. \citealt{Herrmann+2013}), but to facilitate comparison with past studies of IC~1613's inner profile, we focus on (broken) exponentials for our analysis rather than using S{\'e}rsic fits. 

For all the profiles, we used a nonlinear least-squares method for the fits to the data from $12\arcmin$ to $25\arcmin$. For the MS stars, a broken exponential did not represent an improvement over the single exponential fit, while the more complex models had a significantly lower reduced $\chi^{2}$ for the RGB and HB stars.  

The best quality fits for all three stellar tracers are listed in Table~\ref{tab:Parameters} and shown in Figure~\ref{fig:ExpFits}. The bottom panel of the figure shows the surface density profiles after the background is added back. The profiles have been extrapolated in the inner ($r < 12\arcmin$) regions and shown as dotted lines. The lower panels in all the six plots show the fractional residuals left over after the fits.

\subsubsection{Opposite Structures for Younger and Older Stellar Populations}

As stated above, the RGB and HB stars show clear breaks in their exponential profiles. The break radii ($\sim 15\arcmin.8$ and $\sim 17\arcmin.2$, respectively) are consistent, suggesting that they represent the same underlying stellar population. By contrast, the younger MS stars show no break.

Outside of $12\arcmin\ (\sim 2.5 r_h)$ but within the break radii, 
the scale lengths of the populations increase with age: the density of the young MS stars falls off slightly more steeply than the intermediate-age/old RGB stars and much more steeply than the old HB stars.  This is similar to the observations of most dwarf galaxies to date, where the young population is embedded within an older, more extended component \citep{Martinez-Delgado+1999, Aparicio2000, Aparicio+2000,   Hidalgo+2003, Hidalgo+2008, Hidalgo+2009, Vansevicius+2004}. 

This behavior changes in the outer regions beyond the break radii. The young MS stars (well-fit by a single exponential) show the same structure, but the scale lengths of the RGB and HB stars are much smaller than the MS stars, suggesting a transition to a region where their densities fall off more quickly. In detail, the HB profile has the smallest scale length, hinting that the surface density profiles depend monotonically on the age of the population.  A more detailed interpretation is given in Section~\ref{Results}.

\begin{figure}[b!]
    \centering
n    \includegraphics[width = 0.99\columnwidth]{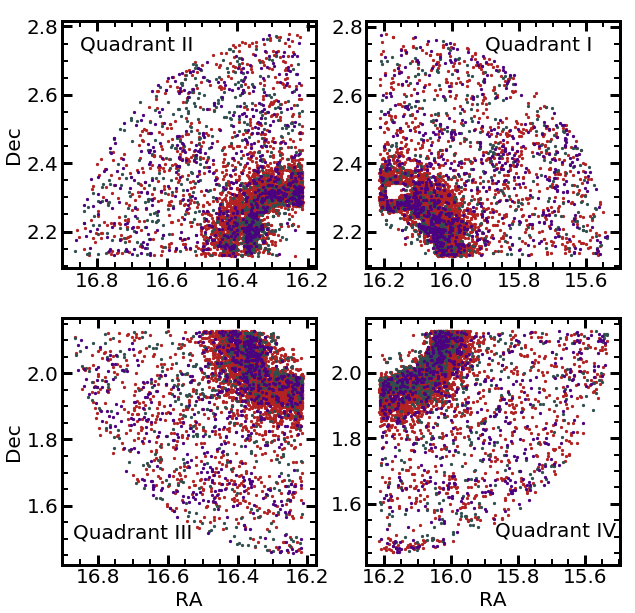}
    \caption{Stellar distribution of RGB (red), HB (purple), and MS (dark green) stars in the four quadrants.}
    \label{fig:Quadrants}
\end{figure}

\begin{figure*}[ht!]
    \centering
    \includegraphics[width = 0.99\textwidth]{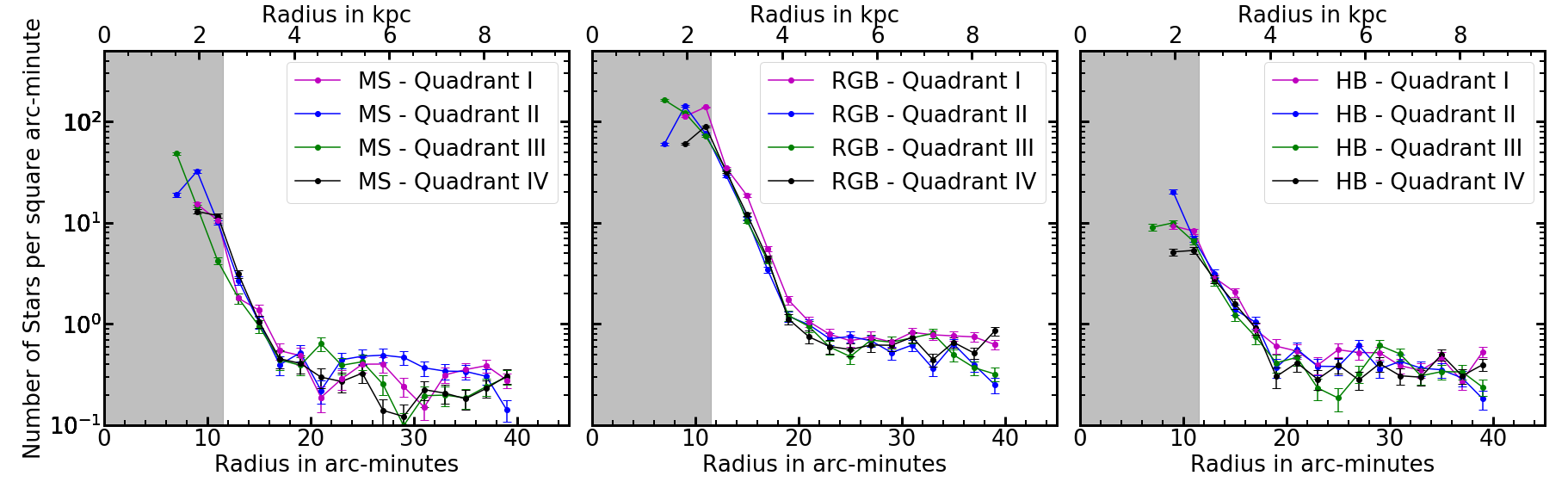}
    \caption{Left: Surface density profiles of MS stars in the four quadrants; Middle: Surface density profiles of RGB stars in the four quadrants; Right: Surface density profiles of HB stars in the four quadrants.}
    \label{fig:Quadrant_Profiles}
\end{figure*}

\subsection{Comparison To Previous Results}

As discussed in Section~\ref{Intro}, previous studies of IC 1613 mostly concentrated on the inner regions of the galaxy.
Here we focus on comparisons to the results of \citet{Bernard+2007}, who also studied the structure of different stellar populations.

\citet{Bernard+2007} constructed radial profiles up to $\sim 15\arcmin$ from the center of IC~1613, and fit exponential functions to populations including MS and RGB stars. For comparison, the scale lengths for these populations are also listed in Table~\ref{tab:Parameters}.

On comparing these scale lengths with the values that we obtained from our fits (see Table~\ref{tab:Parameters}), it is clear that our outer ($r > 12\arcmin$) profiles do not match with their inner ($r < 12\arcmin$) profiles for both RGB and MS stars. This points to a break for both the populations within $12\arcmin$. Combined with the break that we found for the RGB and HB profiles at $\sim 16\arcmin.5$, we conclude that the old population of IC 1613 is made of at least three components, with the profile becoming steeper as we move outwards from the center of the galaxy. The young MS population has at least two components, with the outer component shallower than the inner component.

\subsection{Summarizing the Structure of Stellar Populations}

We list three takeaway points from our investigation of the structure of resolved stellar populations of the dwarf irregular galaxy IC~1613. First, the young MS stars are almost as extended as the old RGB and HB stars, even though the density of young stars is comparatively low in the outer regions of the galaxy. Second, the radial surface density distribution of old stars is steeper in the outer regions compared to the inner regions. Third, the radial surface density distribution of young stars is steeper in the \emph{inner} regions compared to the outer regions: this behavior is opposite to that of the old stars.

We discuss the interpretation of these findings further in Section~\ref{Results}.

\subsection{Looking for Evidence of Accretion} \label{Quadrants}

The presence of an outer break in the structure of the old stellar populations in IC~1613 suggests the possibility of an ``extra" component, which could be identified with an accreted stellar halo. 

Another way to search for evidence of accreted material is to search for evidence of asymmetry in the outer surface density profile. As an initial exploration of this idea, we show a map of the stellar tracers (RGB, HB, and MS) divided into different quadrants is shown in Figure~\ref{fig:Quadrants}.

In each quadrant, the surface density of different stellar tracers is calculated using the method described in Section~\ref{SDP}. The resulting surface density profiles of MS, RGB, and HB stars are given in the left, middle, and right panels of Figure~\ref{fig:Quadrant_Profiles}, respectively. We do not observe any signs of obvious asymmetry in the four profiles for any of the stellar tracers.  Hence we see no evidence, at least from this simple test, for a large-scale asymmetry that could reflect the presence of substantial accretion event(s) in the halo.

\section{Discussion} \label{Results}
Here we discuss the interpretation of the central observational results of the paper: that each studied stellar tracer (MS, RGB, and HB) has a complex, multicomponent structure, with the scale length of the younger stars growing at large radii while the scale length of the older stars shrinks in the outer regions of the galaxy. Young stars are present in small but detectable numbers out to at least $\sim 20$\arcmin~(4.2 kpc $\sim 4 r_h$).

\subsection{Scenarios for Extended Stellar Components}

There are a number of ideas that have been proposed to explain the extended stellar structures in dwarf galaxies. Akin to the classification for more massive galaxies, it is reasonable to separate these explanations into \emph{in situ} scenarios in which the extended structures are formed through internal galactic processes, and scenarios in which the halos are built through the hierarchical accretion of less massive galaxies.

Focusing first on the \emph{in situ} scenarios, there is some evidence from numerical simulations of dwarfs for ``outside-in" star formation. At early times when the gas supply was high, star formation could occur at larger distances from the center, but as gas is consumed, pressure support decreases and star formation is only viable at smaller radii \citep{Stinson+2009}. Scattering in the orbits of old stars and shocks driven by bursts of star formation can also produce more extended distributions of old stars.
\citep{Mashchenko+2008, Stinson+2009, Maxwell+2012}. Using the FIRE simulations, \citet{El-Badry+2016} showed that stellar feedback and bursty star formation can lead to net stellar migration on timescales of a $\sim$ few $\times 10^8$ yr, as well as pushing star-forming gas to larger radii.

Hybrid scenarios where inner old stars are formed \emph{in situ}, but redistributed to large radii through galactic interactions, might also be plausible. For example, 
\citet{Zolotov+2009} use simulations to show that a portion of the inner stellar halos of Milky Way-mass galaxies were indeed formed near the center of the galaxy and displaced to large radii by major mergers. An alternative model is if, akin to the major mergers thought to have formed some early-type galaxies, a low-mass galaxy was formed through
the merger of two lower-mass gas-rich dwarfs. This scenario can lead to a remnant with an extended outer spheroid and an inner star-forming disk \citep{Bekki2008}.

Finally, we consider the standard scenario in which extended structures are built through the accretion of less massive galaxies, as discussed in detail in the Introduction. 
Perhaps the most obvious evidence that this process has occurred would be the the detection of clear streams or tidal features in the halo. More subtle evidence could be the presence of structural features in the radial distribution of stars that could be unambiguously identified with accretion. For example, the stellar density profile of the Milky Way follows a broken power law, with a break at $\sim$ 25 kpc \citep{Juric+2008}. This break radius has been interpreted as a pile-up of stars at their orbital apocenters associated with a single massive accretion event
\citep{Deason+2013, Deason+2018}.

\subsection{Comparison to Observations of IC~1613}

First considering the young MS stars: the presence of such stars at large projected radii ($> 4.2$ kpc; $\sim 4 r_h$) in IC~1613 is puzzling considering that the \ion{H}{1} column density falls to $\sim 10^{16}$ atoms cm$^{-2}$ by $\sim 3.2$ kpc \citep{Hunter+2012}, and at such low densities star formation is unlikely to occur. A recent burst of star formation that pushed gas to larger radii than presently observed could hence explain the presence of young stars at those distances \citep{El-Badry+2016}.

Extending this hypothesis to the older stars, a series of such bursts could also have led to the presence of intermediate-age and old stars at large radii. A possible issue with this scenario is that IC~1613 has an almost constant star formation history \citep{Cole+1999, Skillman+2003, Skillman+2014} with little or no evidence
of distinct bursts. The necessary averaging of the star formation history into bins could smooth out bursts that occurred on timescales $\lesssim 500$~Myr. Nonetheless, it cannot be said that there is specific evidence for the sort of bursts that could have led to large-scale rearrangement of the stellar distribution in IC~1613. Hence this hypothesis seems better-supported for the young stars than for the old stars.

Indeed, the lack of bursts of star formation also presents a problem for any scenario positing a major gas-rich merger, since such a merger is expected to lead to a noticeable elevation in the star formation range \citep{Bekki2008}.

In principle, the breaks in the radial profiles of intermediate-age and old stars could be consistent with either \emph{in situ} outside-in formation (e.g., \citealt{Stinson+2009}) or with accretion of smaller galaxies. The occurrence of breaks in both the RGB and HB stars at a common radius of $\sim 16\arcmin.5$ ($\sim 3.5 kpc$ = 3.4 $r_h$) might be more easily explained in an accretion scenario: accreted galaxies might well have both intermediate-age and old stars that would end up on similar orbits, while similar break radii for a wide range of ages would not seem to be a straightforward prediction of a gradual outside-in formation scenario.

Not all evidence favors the accretion scenario: the lack of evidence for streams or other obvious tidal material is notable. In addition, IC~1613 has no known globular clusters, long considered as luminous tracers of stellar halos \citep{Brodie+2006}, even at large radii (see, e.g., the Local Group search of \citealt{Di+2015}). We did not find any new globular cluster candidates in our data, though without formal artificial cluster tests (planned for a future paper) we cannot quantify these limits on the presence of clusters.

Overall, we conclude that the extended intermediate-age and old components in IC~1613 have observed properties consistent with those expected for accreted stellar halos, but that \emph{in situ} scenarios, such as outside-in star formation, could also explain some of the observed properties of these components, and may separately be required to explain the unusually extended young stars in the galaxy.

\subsection{Future Work}

Detailed numerical simulations could help with the interpretation of this and future, similar data sets. In particular, it would be useful to know whether simulated dwarf galaxies with $M_{\star} \sim 10^7$--$10^9 M_{\odot}$ show evidence for complex radial profiles among their intermediate and older stellar populations, even without a clearly bursty star formation history.  This is one example of the need to better understand the relationship between the outer structure of the galaxy and its \emph{in situ} star formation history. This would better allow one to infer information about the accretion history of the dwarf from the structure of its stellar populations.

On the observational side, spectroscopy of large samples of luminous red giants in the outer regions of IC~1613 would be feasible, and could allow more sensitive searches for evidence of accretion events in phase space than the relatively simple searches in projected physical space that we conducted in this paper.

Looking to future facilities, LSST will extend wide-field observations of dwarf galaxies to 
comparably faint magnitudes over the entire southern sky. From space, JWST and WFIRST will also provide complementary deep and wide-field images that can resolve stars to large radii and offer superior star/galaxy separation to faint magnitudes, allowing one to construct the full star-formation history of IC~1613 and similar dwarfs over a wider range of radii.

\section{Conclusions}
\label{Conclusions}
Using Hyper-Suprime Cam on the Subaru telescope, we have conducted very wide-field deep optical imaging of the Local Group isolated dwarf irregular galaxy IC 1613. By separating the different stellar tracers of MS, RGB, and HB stars from the CMD, we constructed their surface density profiles. Due to completeness constraints, we restricted our analysis to the outer regions ($r > 12\arcmin$), but combined with previous work, we were able to construct profiles over a wider radial range. Our main conclusions are:
\begin{itemize}
    \item[1.] The stellar density profiles of all the three tracers reach background at $\sim 24\arcmin$ (= 5.2 kpc; 5 $r_h$), where the surface brightness is as low as $\sim$ 33.7 mag/arcsec$^{2}$ ($g$-band). 
    \item[2.] The profiles of RGB and HB stars, representing intermediate-age to old stellar populations, have a complex structure with at least three (RGB) and two (HB) components, respectively. The two populations share a common outer ``break" radius of $16\arcmin.5$ (3.5 kpc = $3.4 r_h$) beyond which they steeply decline.
    \item[3.] The young MS stellar profile is composed of two piece-wise components, with the inner component steeper than the outer component. This break radius is interior to $12\arcmin$ but is otherwise not well-determined from our data.
    \item[4.] The young MS stars almost reach the same extent as the old stars, but the outer surface density of young stars is much lower than the older populations, and the density profile of the young stars flattens in the outskirts of the galaxy, again unlike the old populations. This differing behavior between the younger and older stellar populations suggests that they were  assembled in different ways.
    \item[5.] By studying the outer surface density profiles of the stellar tracers in different quadrants, we did not find any obvious (spatial) evidence for nonaxisymmetric material representing an obvious accretion event.
    \end{itemize}

While not definitive, the most straightforward interpretation of these observations is that (i) the extended young population formed from gas that was pushed outward from the center as a result of supernova-driven feedback, and (ii) the outermost intermediate-age and old stars have structural properties more consistent with accretion than \emph{in situ} formation scenarios. 

This latter point is tentative pending additional observational and theoretical studies of the relationship between star formation history and the outer structure of older stars in dwarf galaxies. Such work will bear on our understanding of the assembly of all dwarfs.

\acknowledgments{We acknowledge support from the following NSF grants: AST-1816196 (J.L.C.); AST-1814208 (D.C.); and AST-1813628 (A.H.G.P. and C.T.G.). A.J.R. was supported by National Science Foundation grant AST-1616710, and as a Research Corporation for Science Advancement Cottrell Scholar. J.S. acknowledges support from NSF grant AST-1514763 and the Packard Foundation.  D.J.S. acknowledges support from NSF grants AST1821967, 1821987, 1813708, and 1813466. Research by D.C. is supported by NASA through grants number HST-GO-15426.007-A and HST-GO-HST-GO-15332.004-A from the Space Telescope Science Institute, which is operated by AURA, Inc., under NASA contract NAS 5-26555. We are grateful to Paul Price for his invaluable assistance with data reduction using the LSST pipelines. R.P. thanks Gurtina Besla, Dan Marrone, Don McCarthy, Yumi Choi, Alyson Ford, Deidre Hunter, Robert Lupton, Brenda Namumba, Knut Olsen, D.J. Pisano, and Evan Skillman for their helpful comments on this project.}

\end{document}